\documentclass[epj]{webofc}
\usepackage[varg]{txfonts}   
\wocname{EPJ Web of Conferences}
\woctitle{CONF12}
\newcommand{\mps}{m_{\rm{PS}}}
\newcommand{\fps}{F_{\rm{PS}}}

\newcommand{\mf}{m_{\rm{f}}}

\newcommand{\mpcac}{m_{\rm{PCAC}}}

\newcommand{\ba}{\begin{eqnarray}}
\newcommand{\ea}{\end{eqnarray}}
\newcommand{\be}{\begin{equation}}
\newcommand{\ee}{\end{equation}}

\def\eq#1{Eq.~(\ref{#1})}

\def\fig#1{Fig.~\ref{#1}}

\def\mcO{{\mathcal O}}

\newcommand{\TeV}{\,{\rm TeV}}

\begin{document}
\selectlanguage{english}
\title{Composite Higgs Dynamics on the Lattice}

\author{Claudio Pica\inst{1}\fnsep\thanks{\email{pica@cp3.sdu.dk}} \and
        Vincent Drach\inst{2} \and
        Martin Hansen\inst{1}\and
        Francesco Sannino\inst{1}
}

\institute{CP$^3$-Origins \& Danish IAS,  University of Southern Denmark,  Campusvej 55, DK-5230 Odense M, Denmark 
\and
           Theoretical Physics Department, CERN,  Geneva, Switzerland
}

\abstract{%
We investigate the spectrum of the SU(2) gauge theory with
$N_f$ = 2 flavors of fermions in the fundamental representation, in the
continuum, using lattice simulations.

This model provides a minimal template which has been used for
different strongly coupled extensions of the Standard Model ranging from
composite (Goldstone) Higgs models to intriguing types of dark
matter candidates, such as the SIMPs.
Here we will focus on the composite Goldstone Higgs paradigm, for
which this model provides a minimal UV complete realization in terms
of a new strong sector with fermionic matter.

After introducing the relevant Lattice methods used in our
simulations, we will discuss our numerical results.
We show that this model features a SU(4)/Sp(4) $\sim$ SO(6)/SO(5) flavor
symmetry breaking pattern,  and estimate the value of its
chiral condensate.
Finally, we present our results for the mass spectrum of the lightest
spin one and zero resonances, analogue to the QCD $\rho$, $a_1$, $\sigma$,
$\eta'$, $a_0$ resonances, which are relevant for searches of new, exotic
resonances at the LHC.
\\[.5cm]
{\small CP3-Origins-2016-058 DNRF90}}
\maketitle
\section{Introduction}
\label{intro}
New composite dynamics is often invoked to construct extensions of the Standard Model (SM) physics that can address one or several of the SM shortcomings: for example, composite extensions have been suggested to replace the SM Higgs sector, or to suggest natural dark matter (DM) candidates.   Time-honored examples of  composite dynamics for electroweak symmetry breaking  are Technicolor (TC) \cite{Weinberg:1975gm,Susskind:1978ms} and composite pseudo Nambu-Goldstone Higgs (pNBG) models \cite{Kaplan:1983fs,Kaplan:1983sm}.
 
In TC models the Higgs boson is the lightest scalar excitation of the fermion condensate responsible for electroweak (EW) symmetry breaking~\cite{Sannino:1999qe,Hong:2004td,Dietrich:2005jn,Dietrich:2005wk,Sannino:2009za}. 
The physical Technicolor Higgs mass can be light due to near conformal dynamics \cite{Sannino:1999qe,Dietrich:2006cm} and the interplay between the TC sector and the SM fermions and electroweak gauge bosons~\cite{Foadi:2012bb}. 

In composite Goldstone Higgs models~\cite{Kaplan:1983fs,Kaplan:1983sm}, the new sector has an underlying fundamental dynamics with global flavor symmetry group larger than the one strictly needed to break  the EW symmetry. 
In this case the Higgs state can be identified with one of the additional Goldstone Bosons (GB), and it is therefore naturally light. EW symmetry is broken by radiative corrections, but typically these are not enough and yet another sector is required to induce the correct vacuum alignment for the EW gauge bosons and for the Higgs to acquire the observed mass.

The underlying fundamental theory studied here constitutes the minimal composite template for any natural UV completion that simultaneously embodies both the TC and composite Goldstone Higgs models and it is based on an SU(2) gauge group with two Dirac fermions~\cite{Appelquist:1999dq,Ryttov:2008xe,Galloway:2010bp,Cacciapaglia:2014uja}. 
It is also well known that fermion mass generation constitutes a challenge for any composite dynamics extension. For the present theory an extension that makes use of chiral gauge theories \cite {Appelquist:2000qg,Shi:2015fna} has been put forward recently in \cite{Cacciapaglia:2015yra}, while adding TC-scalars fermions one can nontrivially give mass to all standard model fermions~\cite{Sannino:2016sfx}.
The phenomenology of this model has been studied in detail and it has been showed to be consistent with the present experimental data~\cite{Arbey:2015exa}.

The SU(2) model presented here has also been considered in the context of models of Dark Matter (DM).
In fact, several asymmetric DM candidates were put forward which are stable baryons in TC models~\cite{Nussinov:1985xr,Barr:1990ca} or Goldstone bosons of a new strong sector~\cite{Gudnason:2006ug,Gudnason:2006yj,Ryttov:2008xe,Frandsen:2011kt}.  

Another interesting class of DM models, unrelated to the composite EW scenario,  was recently revived in  \cite{Hochberg:2014dra}. Here an alternative mechanism \cite{Carlson:1992fn,deLaix:1995vi} is employed for achieving the observed DM relic density.  
It uses $3\to2$ number-changing processes that should occur in the dark sector involving strongly interacting massive particles (SIMPs).  
Compared to the WIMP paradigm, where the dark matter particles typically are expected to be around the TeV scale, this model can yield dark matter particles with masses of a few 100 MeVs. In \cite{Hochberg:2014kqa,Hansen:2015yaa} a realization of the SIMP paradigm was introduced in terms of composite theories for which the model investigated here again provides the minimal template.  
Because the energy scale of the SIMP is very light, it is especially relevant to know at which energy scale dark spin-one resonances will appear, or more generally to understand its spectroscopy \cite{Hochberg:2015vrg}. 
Furthermore the new states will modify the scattering at higher energies introducing possible interesting resonant behaviors \cite{Choi:2016hid} and, as it is the case for ordinary QCD, will impact on a number of dark-sector induced physical observables.

Here we present our results for the spectrum of the SU(2) gauge theory with $N_f=2$ flavors of Dirac fermions in the fundamental representation~\cite{Arthur:2016dir,Arthur:2016ozw}, as obtained via numerical lattice simulations.
In particular we discuss out results for the lightest spin-1 and spin-0 resonances, which we obtained after performing both a chiral and a continuum extrapolation. 

The theory has previously been studied on the lattice, and in particular, it has been shown that the expected pattern of spontaneous chiral symmetry breaking is realized~\cite{Lewis:2011zb}. 
Other groups have also investigated the spectrum of this model on the lattice \cite{Hayakawa:2013maa,Amato:2015dqp} concluding that chiral symmetry is broken, although no continuum extrapolation was attempted.
A first estimate, affected by large systematic errors, of the masses of the vector and axial-vector mesons, in units of the pseudoscalar meson decay constant, have been obtained in~\cite{Hietanen:2014xca}. 
The scattering properties of the Goldstone bosons of the theory have also been considered~\cite{Arthur:2014zda}, and the model has furthermore been investigated in the context of possible DM candidates related to the EW scale in~\cite{Hietanen:2013fya,Drach:2015epq}.

\section{A minimal composite Higgs model}
\label{sec-1}

In Nambu-Goldstone composite Higgs models \cite{Kaplan:1983fs,Kaplan:1983sm,Banks:1984gj,Georgi:1984ef,Georgi:1984af,Dugan:1984hq} the Higgs particle is identified with a Nambu-Goldstone boson of a spontaneously broken global flavor symmetry of  a new strongly interacting sector. For a viable realization of this scenario, as long as only electroweak symmetry  is concerned, the symmetry breaking pattern $G_F/H_F$ should be such that the custodial symmetry of the SM is preserved, i.e. $H_F\supset G_{\rm cust}$=SU(2)$_L\times$SU(2)$_R\times$U(1)$_X$, and that one of the Goldstone boson has the correct quantum numbers for the Higgs particle, i.e. belong to the irrep $(2,2)_0$ of $G_{\rm cust}$. 

If we consider UV-complete models in four dimensions featuring fermionic matter, the three minimal cosets are SU(4)$\times$SU(4)/SU(4) for fermions in a complex representation of the gauge group, SU(4)/Sp(4) for fermions in a pseudoreal representation and SU(5)/SO(5) for fermions in a real representation.
The case SU(4)/Sp(4) is realized for a SU(2) technicolor gauge group with $N_f=2$ Dirac fermions considered here, and it therefore represents the minimal realization of a UV-complete pNGB composite Higgs model.
We also note that since SU(4) is locally isomorphic to SO(6) and Sp(4) is locally isomorphic to SO(5), the coset considered here is equivalent to the SO(6)/SO(5) case, sometimes called the next to minimal coset, whose phenomenology has been studied via effective sigma model description, as it is usually the case for pNGB composite Higgs models\footnote{The so-called minimal coset SO(5)/SO(4) for pNGB composite Higgs models lacks a four dimensional UV completion.}.

The Lagrangian of the SU(2) composite Higgs model, in the continuum and for the strong sector in isolation, is simply
\begin{equation}\label{Lagrangian}
\mathcal{L} =  -\frac{1}{4}{F}_{\mu\nu}^a F^{a\mu\nu}
               + \overline{U}(i\gamma^{\mu}D_{\mu}-m)U
               + \overline{D}(i\gamma^{\mu}D_{\mu}-m)D \ ,
\end{equation}
where $U$ and $D$ are the two new fermion fields having a common bare mass $m$, $F_{\mu\nu}^a$ is the field strength, and $D_\mu$ is the covariant derivative.  
The above Lagrangian can be rewritten in terms of the fields 
\begin{equation}
U_{R,L}=\frac{1}{2}(1\pm\gamma^5)U \ ,
~~~~~
D_{R,L}=\frac{1}{2}(1\pm\gamma^5)D \ ,
\end{equation}
and $\widetilde{U}_L= -i\sigma^2C\overline{U}_R^T$, $\widetilde{D}_L = -i\sigma^2C\overline{D}_R^T$ as 
\begin{equation}\label{LagrangianQ}
\mathcal{L} =  -\frac{1}{4}{F}_{\mu\nu}^a F^{a\mu\nu}
               + i\overline{U}\gamma^{\mu}D_{\mu}U
               + i\overline{D}\gamma^{\mu}D_{\mu}D
               + \frac{m}{2}Q^T (-i\sigma^2) C \,EQ + { \frac{m}{2}\left(Q^T(-i\sigma^2)C\,EQ \right)^{\dagger} } \ ,
\end{equation}
where
\begin{equation}
Q = \left( \begin{array}{c} U_L \\ D_L \\ \widetilde{U}_L \\\widetilde{D}_L  \end{array} \right) \ ,
~~~~~~~~
E = \left(\begin{array}{cccc} 0 & 0 & 1 & 0 \\
      0 & 0 & 0 & 1 \\ -1 & 0 & 0 & 0 \\ 0 & -1 & 0 & 0 \end{array}\right) \ ,
\end{equation}
$C$ is the charge conjugation operator acting on Dirac indices, and $-i\sigma^2$ is the antisymmetric tensor acting on color indices. From Eq.(\ref{LagrangianQ}) it is manifest that the mass terms break the global SU(4) symmetry to Sp(4).
In the limit $m\to 0$, the global symmetry is broken spontaneously, as expected and as confirmed by lattice simulations~\cite{Lewis:2011zb}, by the formation of a fermion condensate $\langle \overline{U}U + \overline{D}D \rangle \neq0$.

The choice of a vacuum is not unique, however when considering the theory in isolation, as it is the case in lattice simulations, the physical properties, such as the spectrum and decay constants, do not depend on this choice. 
Once electroweak interactions are added to the model, different vacua of the new strong sector become inequivalent and the physical properties of the model depend on the vacuum alignment.

Following \cite{Cacciapaglia:2014uja}, we embed the new strong sector in the SM by giving the four Weyl fermions the following electroweak charges: $Q_L = (U_L,\, D_L)$ form a SU(2)$_L$ doublet with zero hypercharge; $\widetilde{U}_L$ and $\widetilde{D}_L$ are SU(2)$_L$ singlets with hypercharges $-1/2$ and $+1/2$ respectively. The physically interesting alignments of the condensate are given by $\langle Q^iQ^j  \rangle \propto \Sigma_0$, where $\Sigma_0$ is defined as 
\begin{eqnarray}
\Sigma_0 =\cos \theta\; \Sigma_B + \sin \theta\; \Sigma_H\, ,
~~~~~~~~~
\Sigma_B = \left( \begin{array}{cc}
i \sigma_2 & 0 \\
0 & - i \sigma_2
\end{array} \right)\, ,
~~~~~~~~~
\Sigma_H =\left( \begin{array}{cc}
0 & 1 \\
-1 & 0
\end{array} \right) \ .
\end{eqnarray}
The vacuum $\Sigma_0$ depends on an alignment angle $\theta$: when $\theta=0$, $\Sigma_0=\Sigma_B$ and the fermion condensate does not break electroweak symmetry; on the other end when $\theta=\pi/2$, $\Sigma_0=\Sigma_H$ and the vacuum completely breaks electroweak symmetry. 
These two limits corresponds to the pNGB composite Higgs limit and the Technicolor limit respectively. However the model naturally interpolates between the two. 
The vacuum alignment angle $\theta$ will be determined by the interactions with the electroweak gauge bosons and with the SM fermions, in particular the top quark, not considered so far but which ought to be present in any realistic model for fermion mass generations. 
These interactions with SM fields break the global SU(4) flavor symmetry of the new strong sector via quantum corrections and they will generate a potential for the pNGBs thus fixing the value of $\theta$.
Moreover, additional sources of explicit breaking for the global SU(4) flavor symmetry can be introduced. Therefore the generic case $0<\theta<\pi/2$ is natural. 
The electroweak vev $v$ is related to the scale $f$ of the new strong force\footnote{Here $f$ is the pseudoscalar decay constant of the new strong sector, analogous to $f\pi=93$MeV in QCD.}  via the relation: $v=f\sin\theta$.
The physical Higgs will be a mixture of pNGB and other scalar resonances from the strong sector with the same quantum numbers, i.e., considering only the lightest of such resonances, the pNGB will mix with the $\sigma$-resonance of the new strong sector.

It is worth reminding the other relevant generic features of pNGB composite Higgs models.  In the limit $\theta\to 0$,  the model naturally has couplings to the electroweak gauge boson identical to the ones of the SM: $g_{VVh}=g_{VVh}^{SM}\cos\theta$ and $g_{VVhh}=g_{VVhh}^{SM}\cos 2\theta$; couplings to the SM fermions are also recovered in this limit, although the specific details are model-dependent: in our case $g_{h\bar{f}f}=g_{h\bar{f}f}^{SM}\cos\theta$; finally the Peskin-Takeuchi S-parameter becomes small: $S\propto \sin^2\theta$. 

As studied in \cite{Arbey:2015exa}, experimental data sets constraints on the allowed values of the alignment angle: a typical upper limit on the alignment angle is $\sin\theta<0.2$.

\section{Lattice results}
\label{sec-2}

Here we will study the spectrum of the strong sector in isolation, via lattice simulations. 
We discretize the SU(2) gauge theory with two Dirac fermions in the fundamental
representation of \eq{Lagrangian} using the (unimproved) Wilson action for the two mass-degenerate Dirac fermions and the Wilson plaquette action for the gauge fields.
The numerical simulations have been performed using an updated version of the HiRep code~\cite{DelDebbio:2008zf}.
The fermionic part of the action reads:
 \be
S_F =  \sum_x\overline{\psi}(x)(4+am_0)\psi(x)
    - \frac{1}{2}\sum_{x,\mu}\left(\overline{\psi}(x)(1-\gamma_\mu)U_\mu(x)\psi(x+\hat\mu)
   +\overline{\psi}(x-\hat\mu)(1+\gamma_\mu)U_\mu^\dagger(x)\psi(x)\right) \,,
\ee
where $U_\mu$ is the gauge field,  $\psi$ is the doublet of $U$ and $D$ fermions, and $am_0$ is
the $2\times2$ diagonal mass matrix proportional to the identity.

Our simulation are performed at four values of the inverse lattice gauge coupling $\beta=2N/g^2$, for a number of different fermion masses and on several lattice volumes. This is needed in order to perform the required extrapolations to the chiral limit and infinite volume and to give an estimate of the systematic errors stemming from such extrapolations. We used the four different lattice spacings, corresponding to four different UV cutoffs, to extrapolate our results to the continuum limit. The full set of parameters used for the analysis and a detailed description of the procedures involved in the extrapolations are given in \cite{Arthur:2016dir,Arthur:2016ozw}. In particular we refer to these references for the details about the extraction of the resonance masses from numerical lattice data.

Here we summarize the main steps in the analysis and present our results.

\subsection{Scale setting and non-perturbative renormalization}\label{sub-1}
Recent developments stemming from the so-called ``Wilson-flow''~\cite{Luscher:2010iy}, led to the introduction of two different scale-setting observables, known as $t_0$~\cite{Luscher:2010iy} and $w_0$~\cite{Borsanyi:2012zs}, which can be measured much more precisely then other observables used in the past for the same purpose. In this work we used $w_0$, although similar results are obtained from $t_0$.

\begin{figure}[t]
  \centering
\begin{minipage}{.48\textwidth}
  \centering
 \includegraphics[width=\linewidth]{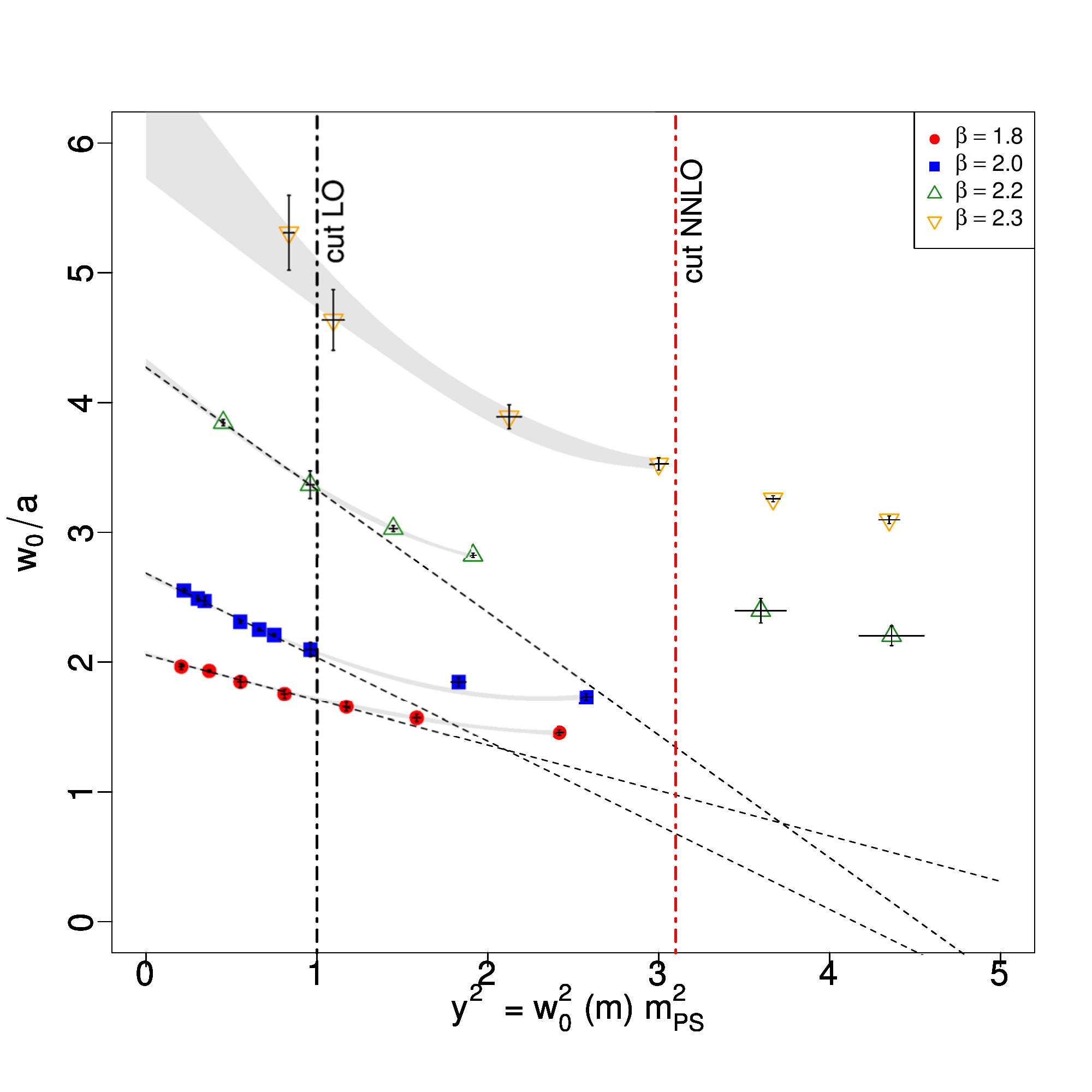}
\end{minipage}%
\hspace*{0.5cm}\begin{minipage}{.48\textwidth}
\centering
  \includegraphics[width=\linewidth]{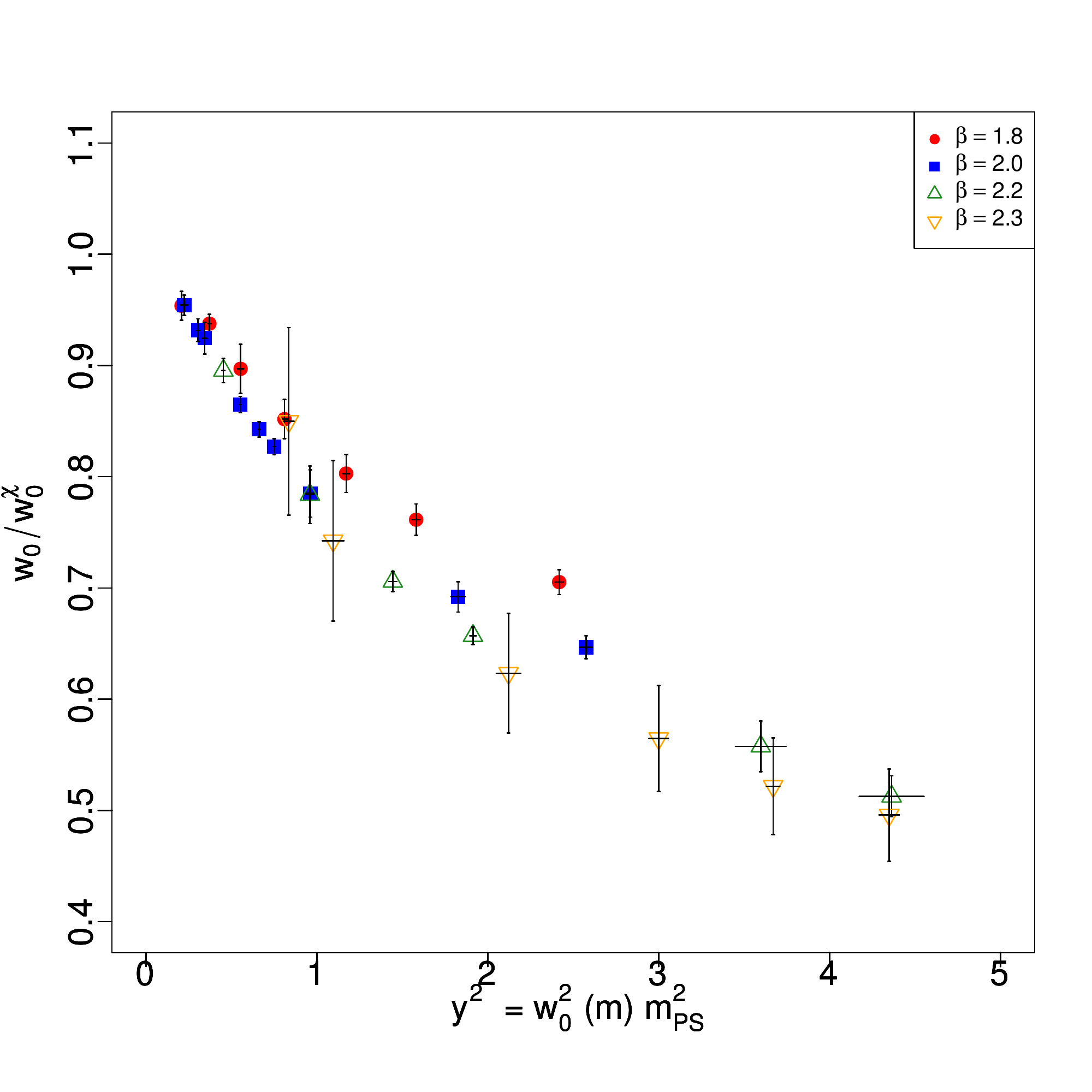}
\end{minipage}
\caption{Chiral behavior of $w_0$ as a function of $y^2$ in unit of the lattice spacing (left panel) and in unit of $w_0^{\chi}$ (right panel) for $W_{\rm{ref}}=1$. The data at four lattice spacings are displayed. \label{fig:w0_vs_y2} }
\end{figure}

In Fig.~\ref{fig:w0_vs_y2} (left panel) we show our results for $w_0/a$ for the four lattice spacings considered in this study as a function of $y=w_0(\mpcac)\,\mps$, where $\mpcac$ is the current quark mass and $\mps$ is the pNGB mass. 
For all the points in  Fig.~\ref{fig:w0_vs_y2} we have $\mps\,L > 5.5$, which leads to negligible finite volume effects for this observable, as confirmed numerically.

In order to extrapolate to the chiral limit, we used the NNLO expansion in terms of $\mps^2$ which reads~\cite{Bar:2013ora}:
\be
w_0(\mps^2) = w^\chi_0 \left( 1 + k_1 \frac{m_{\rm{PS}}^2}{(4\pi F)^2} +  k_2
\frac{m_{\rm{PS}}^4}{(4\pi F)^4} \log \frac{m_{\rm{PS}}^2}{\mu^2} \right)\,,\label{eq:w0}
\ee
where $F$ is the pseudoscalar decay constant and $k_1$, $k_2$ are dimensionless low energy constants. Note that the chiral logarithm enters only at NNLO. Rewriting \eq{eq:w0} above, we fitted our data at each $\beta$ with the following function :
\be
w_0(\mps^2)= w^\chi_0 \left( 1 + A y^2 + B y^4\log y^2\right)\,,
\ee
where $A$, $B$ and $w^\chi_0$ are free parameters (we also set $w^\chi_0 \mu = 1$). This functional form describes well our numerical data, and from it we can determine the lattice spacing in the chiral limit from $w^\chi_0$.

In the right panel of \fig{fig:w0_vs_y2} we show $w_0/w_0^{\chi}$  for all four lattice spacings. The  deviation from a universal curve of such a quantity is a measure of lattice discretization errors. As can be seen, these are small in the $w_0$ observable for our three finest lattice spacings corresponding to larger $\beta$ values. 

To properly renormalize the pseudoscalar decay constant $\fps$ and of the quark mass $\mpcac$, we determine the non-perturbative renormalisation constants of the isovector vector (V), axial (A), and pseudoscalar (P) bilinear operators. 
We use the RI'-MOM scheme (regularization invariant momentum scheme) as in~\cite{Martinelli:1994ty}, which is defined imposing renormalization conditions on the fermion propagator and amputated Green's function for fermion bilinear operators in the chiral limit and at a given reference momentum. We verified that our conclusions are not affected by this choice of reference momentum, within the accuracy of this study.

\begin{figure}[t!]
  \centering
  \includegraphics[width=.48\textwidth]{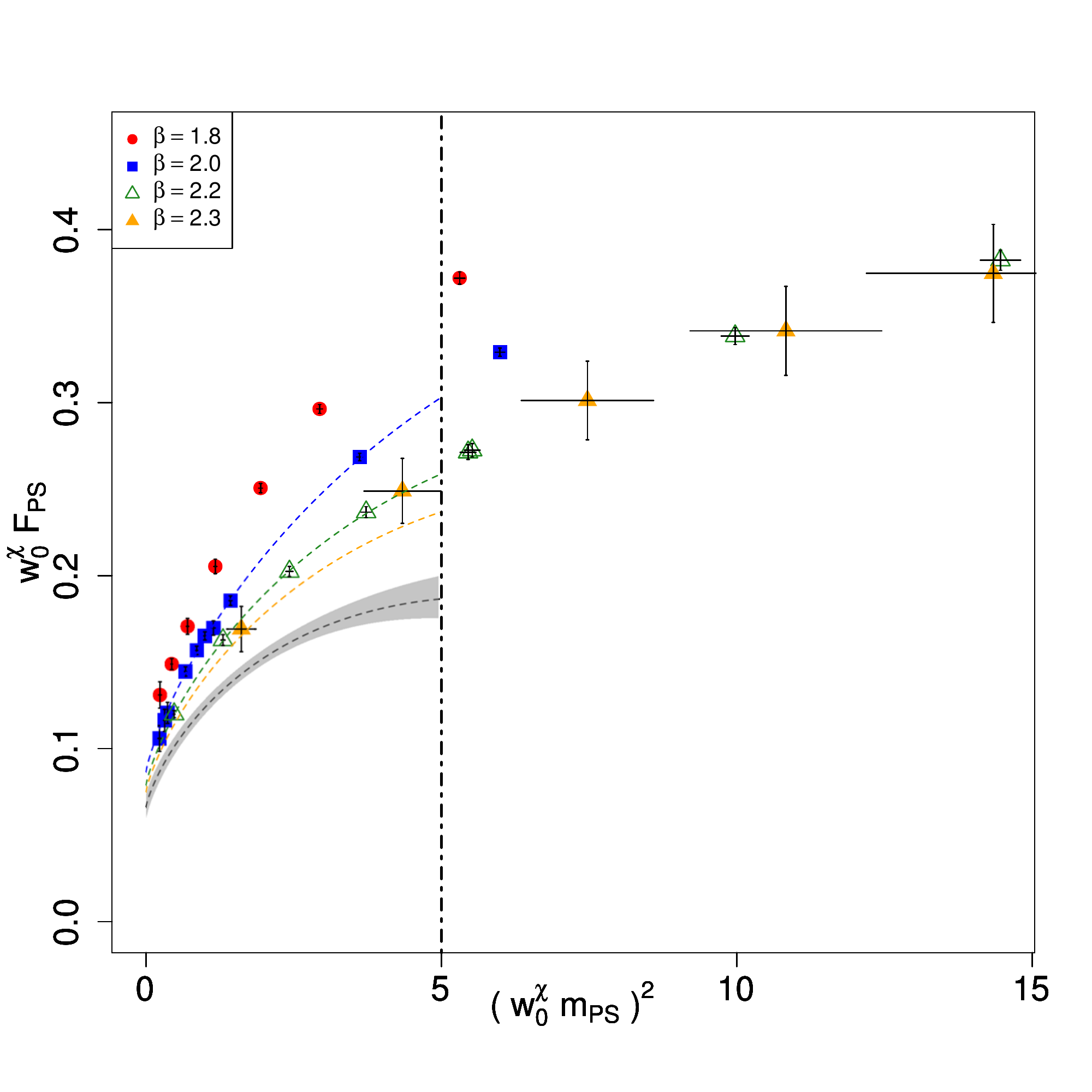}
\hfill
  \includegraphics[width=.48\textwidth]{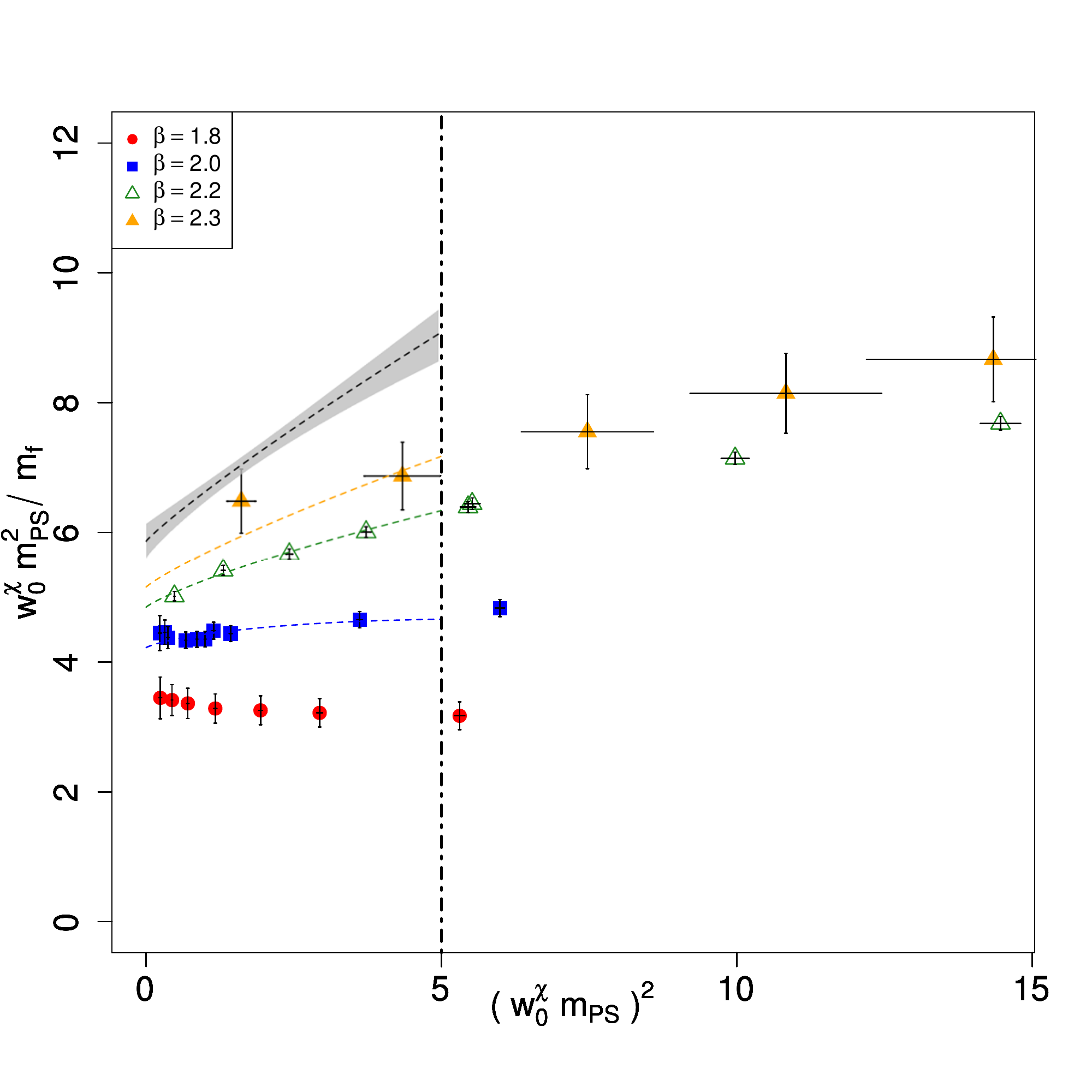}
  \caption{$\fps$ versus $\mps^2$ (left panel) and $\mps^2/\mf$ versus $\mps^2$ (right panel) for the four lattice spacings used in this work. The curves correspond to the best fit parameters to \eq{eq:NLO_mps_fps} and (\ref{eq:NLO_mps_fps2}). The black dashed curve indicate the continuum results.}
  \label{fig:w0mps2_over_mf_vs_mps}
\end{figure}

\subsection{Goldstone sector: $\mps$ and $\fps$}\label{sub-2}
Here we describe the procedure used to extract the relevant low energy constants in the chiral limit from our numerical data obtained at a finite cutoff and with a finite quark mass. We consider the expressions at next-to-leading order chiral perturbation theory in the continuum, and at infinite volume, for the Goldstone boson mass $\mps$ and its decay constant $\fps$ which read~\cite{Bijnens:2009qm}:
\be\label{eq:NLO_cont_mps_fps_xtilde}
\frac{\mps^2}{\mf} = 2 B  \left[ 1 +\frac{3}{4} \tilde{x} \log{\frac{\mps^2}{\mu^2}} +   b_M  \tilde{x} +\mcO(\tilde{x}^2) \right]\,,\,\, 
\fps = F\left[ 1 - \tilde{x}\log{\frac{\mps^2}{\mu^2}} +  b_F \tilde{x}  +\mcO(\tilde{x}^2) \right]\,,
\ee
where $\tilde{x} = \frac{\mps^2}{(4\pi F)^2}$, $\mf(p^2)= Z_A/Z_P(p^2)\mpcac$ is the renormalized fermion mass at the given scale reference scale and $\fps =\fps^{\rm (bare)} Z_A$ the renormalized pseudoscalar decay constant.
In the conventions of \cite{Bijnens:2009qm}, the condensate is given by $\Sigma \equiv - 2 B F^2$. 
The use of the infinite volume expressions is justified given the values of $\fps L$ and $\mps L$, with $L$ the lattice size, for our numerical simulations.
To take into account cutoff effects we use the following parametrization: 
\ba\label{eq:NLO_mps_fps}
\frac{\mps^2}{\mf} &=& 2 B  \left[ 1 - a_M \tilde{x} \log{\frac{\mps^2}{\mu^2}} +   b_M  \tilde{x}   +\delta_M \frac{a}{w^{\chi}_0} +\gamma_M \mps^2 \frac{a}{w^{\chi}_0}  \right]\,, \\
\fps &=& F\left[ 1 -a_F \tilde{x}\log{\frac{\mps^2}{\mu^2}} +  b_F \tilde{x}  +\delta_F \frac{a}{w^{\chi}_0} +\gamma_F \mps^2 \frac{a}{w^{\chi}_0}  \right]\label{eq:NLO_mps_fps2}\,.
\ea
Here the new fitting parameters $\delta_{M,F}$ and $\gamma_{M,F}$ control the discretization effects. Note also that the two coefficients $a_{F,M}$ are fixed in the continuum, but here we consider them as free parameters at finite lattice spacing. This parametrization is inspired by Wilson chiral perturbation theory.

We show in \fig{fig:w0mps2_over_mf_vs_mps} the results of our chiral and continuum extrapolations obtained from \eq{eq:NLO_mps_fps} and (\ref{eq:NLO_mps_fps2}). Due to large cutoff effect, the data at our coarsest lattice spacing is not well described by the parametrization used, and it was not included in the final fit.

Our final estimates for the chiral parameters are $w_0^{\chi} B=2.88(15)(17)$ and $w_0^{\chi} F=0.078(4)(12)$. The systematic error was obtained by comparing the results from the above procedure, with a different extrapolation strategy, namely performing a chiral extrapolation at fixed lattice spacing first, followed by a linear continuum extrapolation of the parameters.
%
The value of the condensate then reads $\Sigma^{1/3}/F = 4.19(26)$ (statistical and systematical errors have been combined).

\begin{figure}[t!]
  \centering
  \includegraphics[width=.48\textwidth]{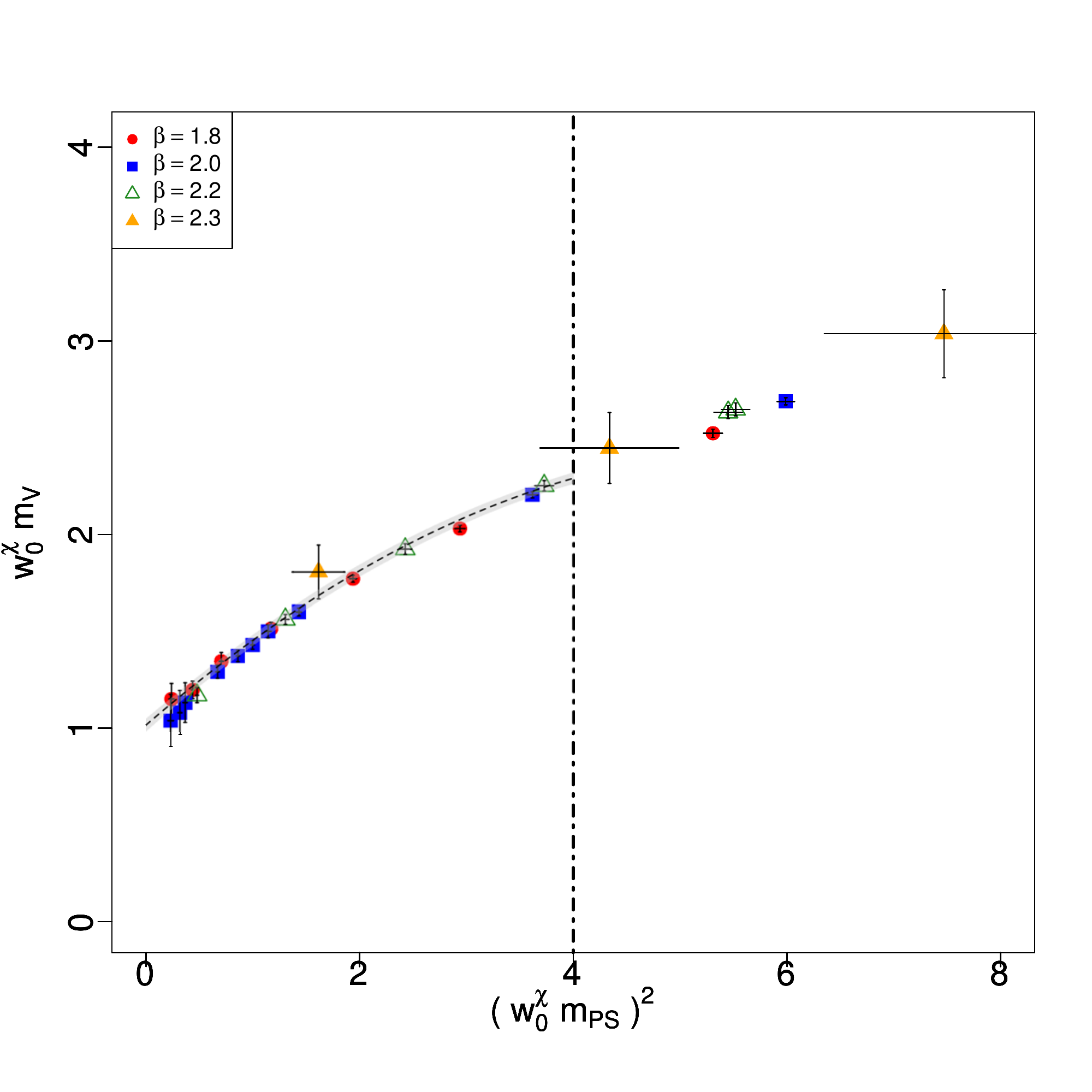}
\hfill
  \includegraphics[width=.48\textwidth]{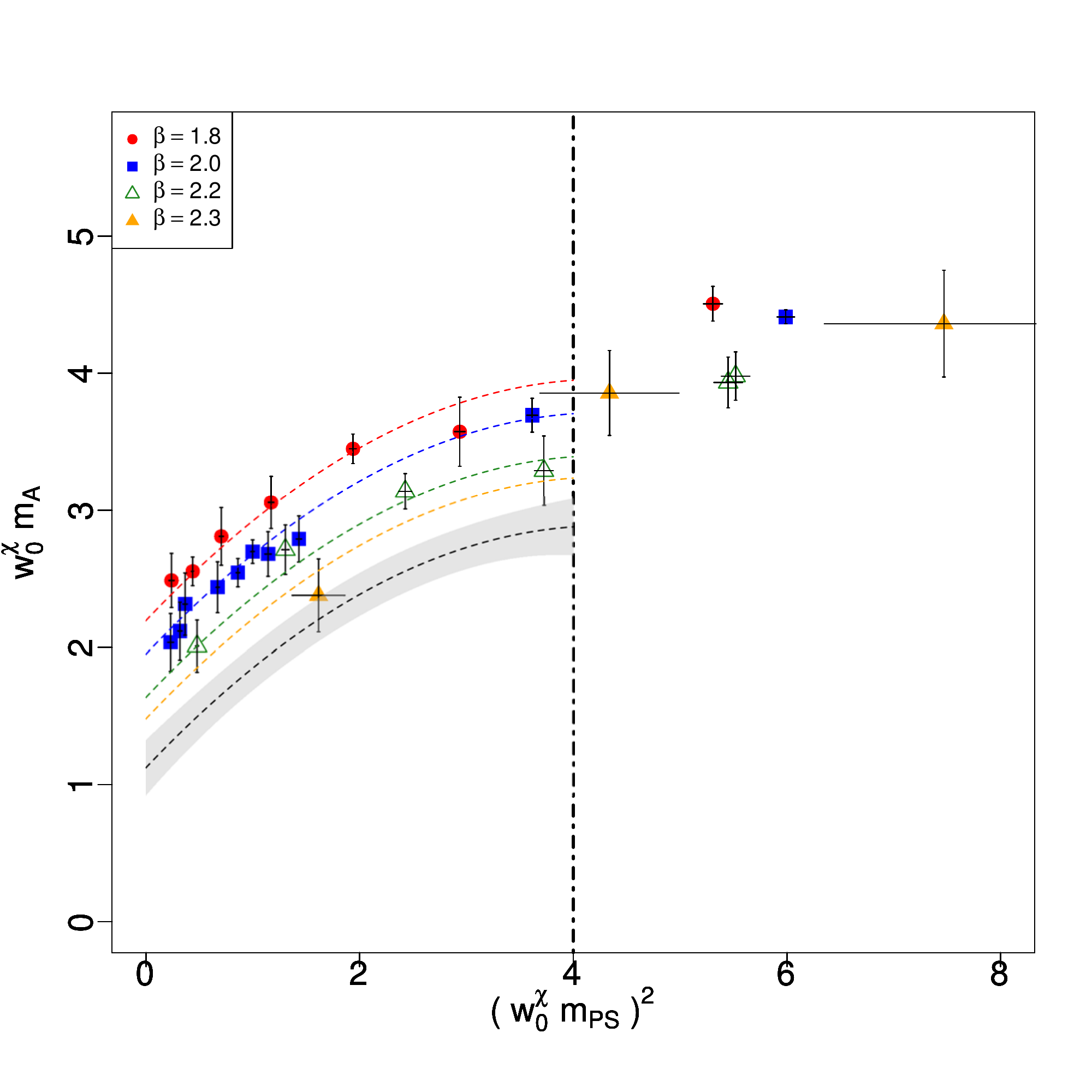}
  \caption{Vector $m_V$ (left panel) and axial-vector $m_A$ (right panel) meson masses versus $\mps^2$. The curves correspond to the best fit parameters to \eq{eq:heavyans}. The black dashed curve indicate the continuum results.}
  \label{fig:w0mV}
\end{figure}

\begin{figure}[p!]
  \centering
  \includegraphics[width=.48\textwidth]{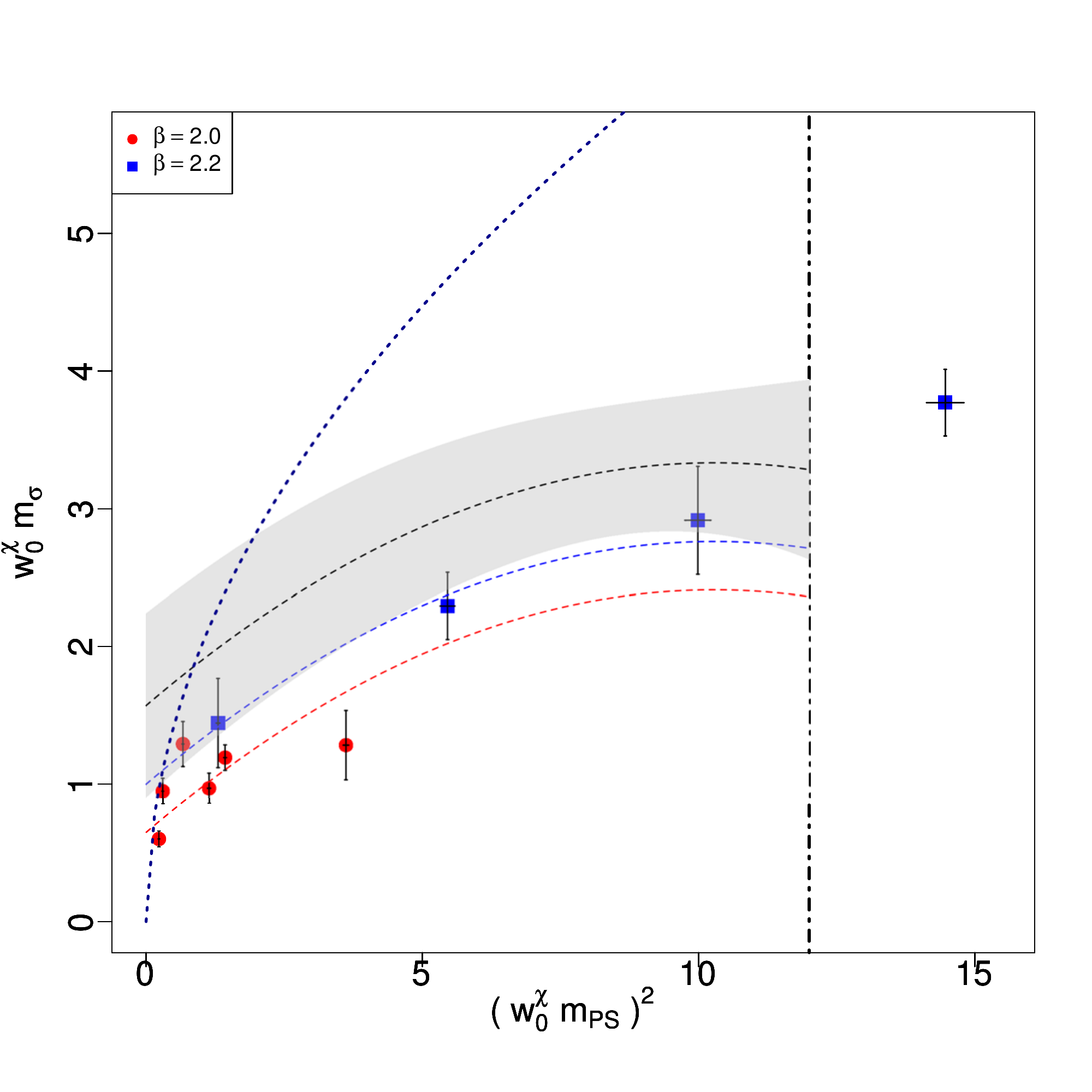}
\hfill
  \includegraphics[width=.48\textwidth]{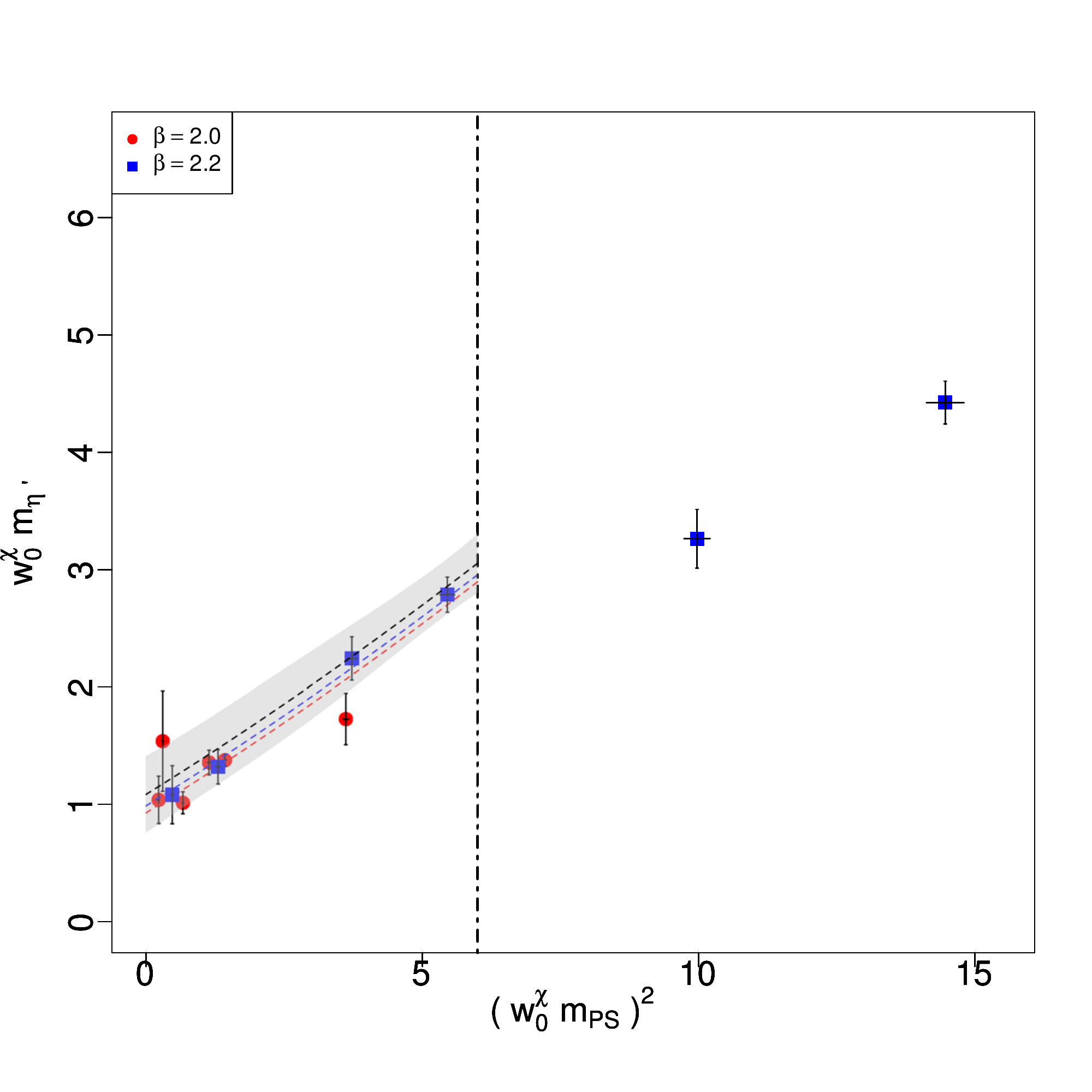}
  \caption{Mass of spin-0 isosinglet resonances $m_\sigma$ (left panel) and $m_{\eta'}$ (right panel) versus $\mps^2$. The curves correspond to the best fit parameters to \eq{eq:heavyans}. The black dashed curve indicate the continuum results.}
  \label{fig:w0ms}
\end{figure}
\begin{figure}[p!]
  \centering
  \includegraphics[width=.48\textwidth]{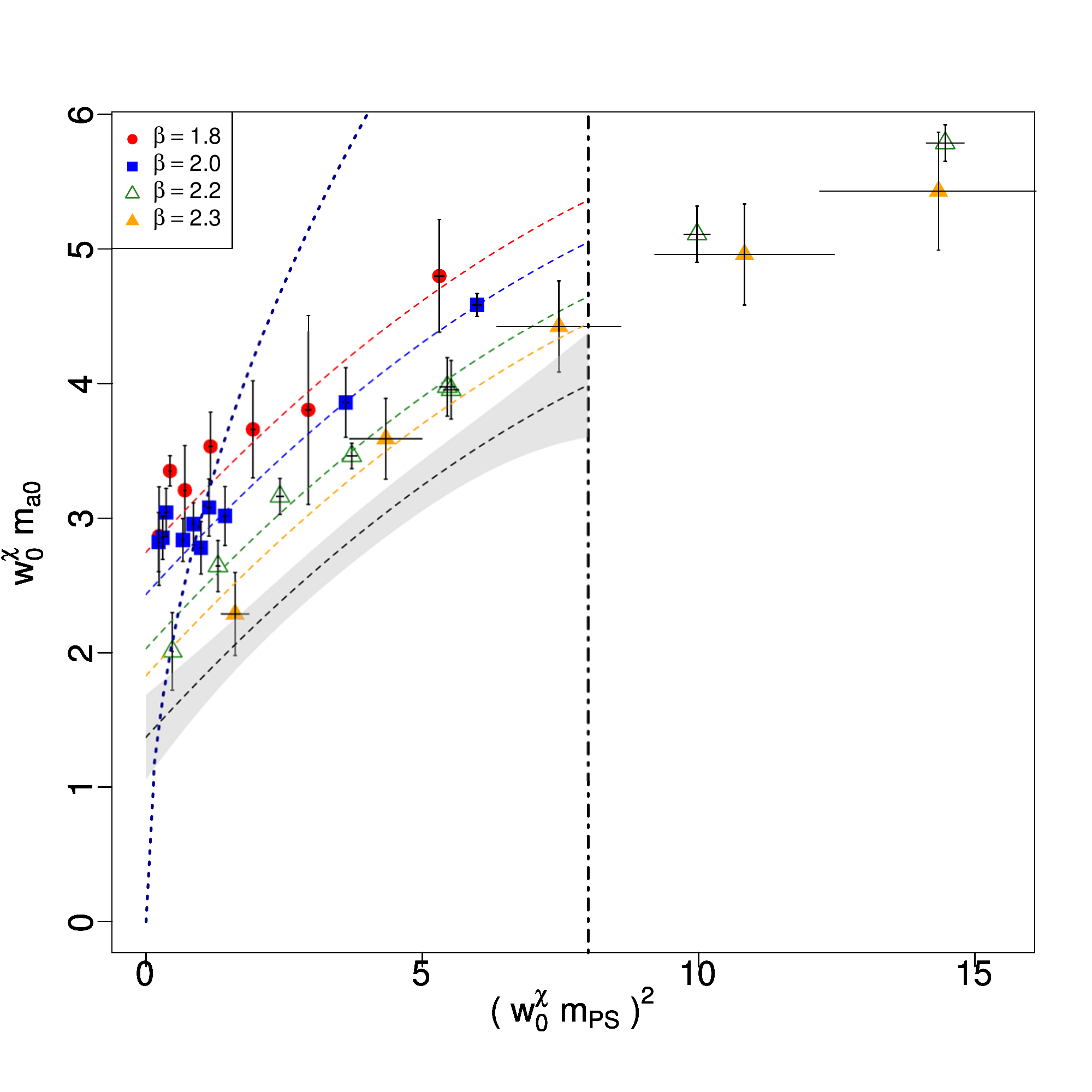}
  \caption{Mass of spin-0 isotriplet $m_{a_0}$ resonance versus $\mps^2$. The curves correspond to the best fit parameters to \eq{eq:heavyans}. The black dashed curve indicate the continuum results.}
  \label{fig:w0ma0}
\end{figure}
\subsection{Spin-1 resonances}\label{sub-3}

We now consider the isotriplet vector and axial-vector mesons. In order to perform the required chiral and continuum extrapolations we use a similar strategy to the one used in the previous section. We perform a combined chiral and continuum extrapolation by fitting our data to the following ansatz:
\be
w_0^\chi m_X = w_0^\chi m_X^\chi + A  (w_0^\chi \mps)^2 + B  (w_0^\chi\mps)^4 + C \frac{a}{w_0}\, .\label{eq:heavyans}
\ee
The results of the fit are shown in \fig{fig:w0mV} for the vector (left panel) and the axial vector (right panel) mesons.

For the vector meson the fit describes our data well and the observed cutoff effects are small. We find $w_0^{\chi} m^\chi_V = 1.01(3)$ with a $\chi^2/\textrm{ndof}=23/16$.
For the mass of the axial-vector meson, our data is more noisy already at the level of the effective masses and we therefore have larger systematic uncertainties.  The ansatz \eq{eq:heavyans} fits the data well, within large errors, and the resulting value for the mass is: $w_0^\chi m^\chi_A =1.1(1)$ with $\chi^2/\textrm{ndof}=20/16$.
In units of $\fps$ we have $m_V/\fps\sim 13.1(2.2)$ and $m_A/\fps\sim 14.5(3.6) $.
These values are higher than the one from QCD, and we also note that the large error comes mainly from the continuum extrapolation of $\fps$ and $m_A$.

\subsection{Spin-0 resonances}\label{sub-4}

We performed a first benchmark computation of scalar resonances, corresponding to the $\sigma$, $\eta'$ and $a_0$ resonances in QCD. In the case of isosinglet states, the $\sigma$ and $\eta'$, the numerical estimation is very challenging due to the presence of disconnected contributions in the two-point functions used to estimate the mass of the states.
For this reason we are not able to obtain a signal for all our ensembles of parameters. We present here our first estimate, which is affected by large systematic errors.

Also for scalar resonances we use \eq{eq:heavyans} for a combined chiral and continuum extrapolation.

We show in \fig{fig:w0ms} our numerical data and the extrapolation for the isosinglet scalar states, and in \fig{fig:w0ma0} the results for the $a_0$ resonance.

In the case of the $a_0$, we have checked that finite volume effects are not significant on three different volumes ($L/a=16,24,32$). Since the estimate of the mass of the $a_0$ does not require the estimate of any disconnected loops contribution, we are able to obtain a signal on  all data sets and thus include four lattice spacings in the extrapolation.
However since we observe that some of our data points lie above the 3 Goldstone boson mass threshold, we exclude these data points from the fit. 

Our final estimates for the scalar meson masses are:  $w^{\chi}_0m_{a_0}= 1.3(3)$, $w^{\chi}_0m_\sigma=1.5(6)$ and $w^{\chi}_0 m_{\eta'} = 1.0(3)$. 
In units of $\fps$ these are: $m_{a_0}/\fps= 16.7(4.9)$, $m_\sigma/\fps=19.2(10.8)$ and $m_{\eta'}/\fps = 12.8(4.7)$.

\begin{figure}[t!]
  \centering
  \includegraphics[width=.48\textwidth]{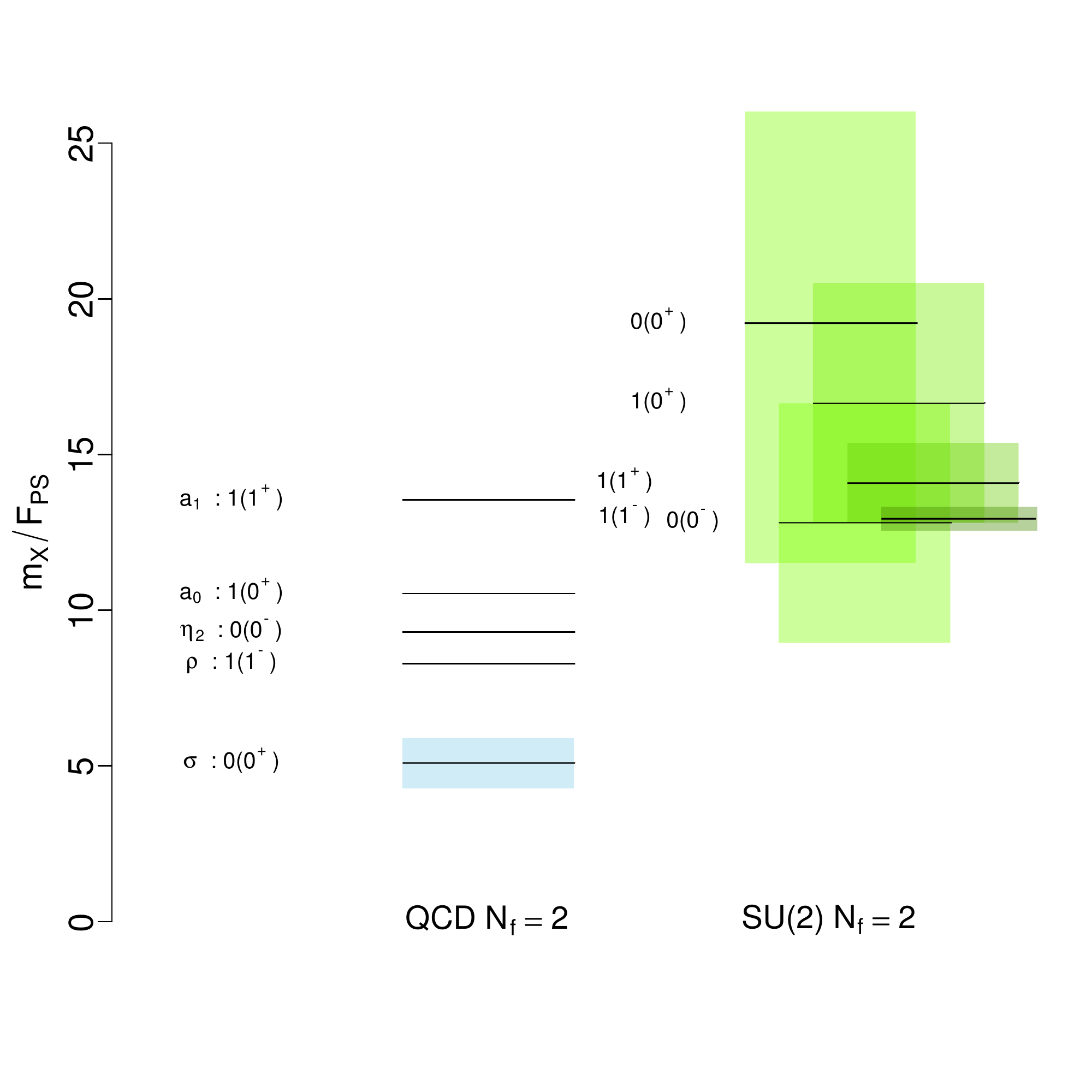}
  \caption{Comparison of spectrum of QCD ($N_f=2$) with our current
    results. We use the notation $I(J^P)$ to label states. The QCD result are
    taken from experiments (at the physical value of the pion mass)
    except for the $\eta'$ where we took the central value of a
    $N_f=2$ lattice calculation (denoted
    $\eta_2$)\cite{Jansen:2008wv}. The error on the QCD sigma pole mass
    is shown by a light blue band. Note that while in two flavor QCD
    every $I=1$ state is triply degenerate, in our case $I=1$ states correspond to five degenerate states.}
  \label{fig:spectrum}
\end{figure}

\section{Conclusions}

We analyzed the  SU(2) gauge theory with $N_f = 2$ fermions in the fundamental representation using lattice techniques. This model has been used as a minimal template for a UV complete pNGB composite Higgs model, compatible with current experimental limits. We presented here the first results for the low lying spectrum and low energy constants of the model, including both chiral and continuum extrapolations. 

Our results include a detailed analysis of the Goldstone sector, which lead to the determination of $\fps$ used in the model to set the new strong force scale via the relation: $246\ {\rm GeV}=\fps\sin\theta$, where $\theta$ is the vacuum alignment angle of the model.
The spectrum for both the lightest spin-1 and spin-0 resonances was also determined. 
Our results, in terms of $\fps$, are: $ m_V/\fps=13.1(2.2)$, $ m_A/\fps=14.5(3.6)$, $m_{a_0}/\fps= 16.7(4.9)$, $m_\sigma/\fps=19.2(10.8)$ and $m_{\eta'}/\fps = 12.8(4.7)$.

Although we take great care to estimate all sources of systematic errors, in some cases the present numerical data is too limited for a reliable estimation and therefore, in these cases, systematic errors might be underestimated. In particular this is  the case for the isosinglet scalar resonances, which are notoriously very hard to estimate on the lattice.

A summary of our results, as compared to $N_f=2$ QCD, are shown in \fig{fig:spectrum}. Taken at face value, our present results indicate a spectrum which is quite different from the QCD one, featuring heavier resonances.
As an example our results predicts new vector resonances of mass:
\be
m_V = \frac{3.2(5)}{\sin \theta}~\TeV,\quad\text{and}\quad m_A = \frac{3.6(9)}{\sin \theta}~\TeV\,,
\ee
which are beyond the present LHC constraints, even in the Technicolor limit \cite{Franzosi:2015zra} where $\theta=\pi/2$.
In the pNGB limit, for $\sin\theta<0.2$, these resonances seem beyond the reach of LHC experiments. Therefore in this minimal UV complete realization of the pNGB composite Higgs model, resonances from the new strong sector might not be detected at experiments, in contrast to what usually assumed for pNGB composite Higgs models.

Our results are still affected by large systematic errors, mainly due to the chiral and continuum extrapolations required to obtain predictions for phenomenologically relevant models. In the future we plan to increase the accuracy of the our results, in particular by improving the quality of our continuum extrapolations and the precision of the mass measurements for isosinglet scalar mesons.

This work was supported by the Danish National Research Foundation DNRF:90 grant and by a Lundbeck Foundation Fellowship grant. The computing facilities were provided by the Danish Centre for Scientific Computing and the DeIC national HPC center at SDU.

%
\bibliography{Bib}

\end{document}